\documentclass{PoS}

\title{Testing the magnetic field models of disk galaxies with the SKA}

\ShortTitle{Testing the magnetic field models of galaxies with the
SKA}

\author{\speaker{T.G.~Arshakian},$^a$ R.~Stepanov$^{b}$, R.~Beck$^{a}$, P.~Frick$^{b}$ and M.~Krause$^{a}$\\
\llap{$^a$}Max-Planck-Institut f\"ur Radioastronomie, Auf dem H\"ugel 69, 53121 Bonn, Germany\\
\llap{$^b$}Institute of Continuous Media Mechanics, Korolyov str.~1,
  614061 Perm, Russia\\
E-mail: \email{tigar@mpifr-bonn.mpg.de}, \email{rodion@icmm.ru},
\email{rbeck@mpifr-bonn.mpg.de}, \email{frick@icmm.ru},
\email{mkrause@mpifr-bonn.mpg.de}}

\abstract{The future new-generation radio telescope SKA (Square
Kilometre Array) and its precursors will provide a rapidly growing
number of polarized radio sources. Hundred and thousands polarized
background sources can be measured towards nearby galaxies thus
allowing their detailed magnetic field mapping by means of Faraday
rotation measures (RM). We aim to estimate the required density of
the background polarized sources detected with the SKA for reliable
\emph{recognition} and \emph{reconstruction} of the magnetic field
structure in nearby spiral galaxies. We construct a galaxy model
which includes the ionized gas and magnetic field patterns of
different azimuthal symmetry (axisymmetric (ASS), bisymmetric (BSS)
and quadrisymmetric spiral (QSS), and superpositions) plus a halo
magnetic field. RM fluctuations with a Kolmogorov spectrum due to
turbulent fields and/or fluctuations in ionized gas density are
superimposed. \emph{Recognition} of magnetic structures is possible
from RM towards background sources behind
galaxies or a continuous RM map obtained from the diffuse polarized
emission from the galaxy itself. Under favourite conditions, about a
few dozens of polarized sources are sufficient for a reliable
recognition. \emph{Reconstruction} of the field structure without
precognition becomes possible for a large number of background
sources. A reliable reconstruction of the field structure needs at
least 20 RM values on a cut along the projected minor axis which
translates to $\approx1200$ sources towards the galaxy. Radio
telescopes operating at low frequencies (LOFAR, ASKAP and the
low-frequency SKA array) may also be useful instruments for field
recognition or reconstruction with the help of RM, if background
sources are still significantly polarized at low frequencies. This
work is a part of the science simulations of cosmic magnetism in the
frame of the SKADS (SKA Design Studies).}

\FullConference{From Planets to Dark Energy: the Modern Radio Universe\\
                 October 1-5 2007\\
                 The University of Manchester, UK}

\begin{document}

\section{Introduction}
Faraday rotation measures (RM) in galaxies are generated by regular
fields of the galaxy plus its ionized gas, both of which extend to
large galactic radii (Fig.~\ref{fig:m31}). RM towards polarized
background sources can trace regular magnetic fields in these
galaxies out to even larger distances, however, with the sensitivity
of present-day radio telescopes, the number density of polarized
background sources is only a few sources per solid angle of a square
degree, so that only M~31 and the LMC could be investigated so far
\cite{han98,gaensler05}.

Future high-sensitivity radio facilities will observe polarized
intensity and RM for a huge number of faint radio sources, thus
providing the high density background of polarized point sources.
This opens the possibility to study in detail the large-scale
patterns of magnetic fields and their superpositions thus allowing
the dynamo theory for field amplification and its ordering to be
tested. A major step towards a better understanding of galactic
magnetism will be achieved by the Square Kilometre Array (SKA,
www.skatelescope.org) and its pathfinders.
%, the Allen Telescope
%Array (ATA) in the US, the Low Frequency Array (LOFAR) in Europe,
%the MeerKAT in South Africa, and the Australia SKA Pathfinder
%(ASKAP). The SKA will be a new-generation telescope with a square
%kilometre collecting area, a frequency range of 70~MHz to 25~GHz
%with continuous frequency coverage, a bandwidth of at least 25\%, a
%field of view of at least 1 square degree at 1.4~GHz, and angular
%resolution of better than 1~arcsecond at 1.4~GHz. The SKA is planned
%to consist of three separate arrays: a phased array for low
%frequencies (70--300~MHz), a phased array for medium frequencies
%(300--1000~MHz), and an array of single-dish antennas for high
%frequencies (1--25~GHz).

\section{Models of RM towards foreground galaxy}
\begin{figure*}
\begin{center}
  \includegraphics[width=0.7\textwidth]{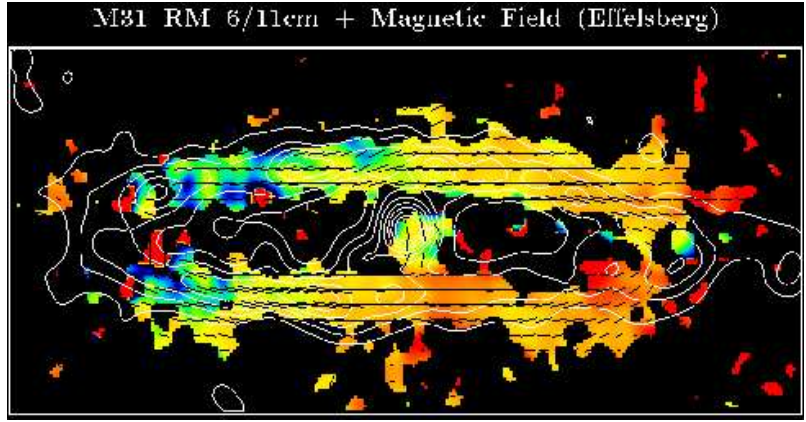}\\
  \caption{M\,31: total emission (contours) and polarized emission
  (B-vectors) at $\lambda $6.2 cm superimposed with the distribution
  of the Faraday rotation measure from diffuse radio emission \cite{berk03}.}
  \label{fig:m31}
\end{center}
\end{figure*}

\begin{figure*}
\begin{center}
\includegraphics[width=0.8\textwidth]{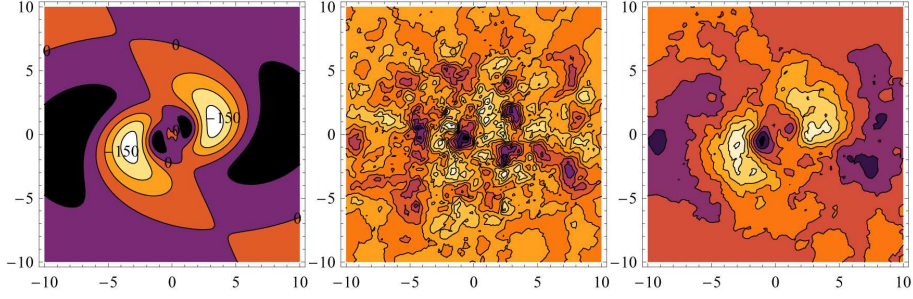}
\caption{'Template' RM map (in rad m$^{-2}$) for an inclination
angle of $i=10^o$ generated for a pure BSS field pattern, for
$B=5\,\mu$G and $n_0 $=0.03~cm$^{-3}$ (left), a random turbulent field
with r.m.s. of $30$~rad m$^{-2}$ (middle) and their superposition, 
the 'observed' RM map (right).}
\label{fig:rm_model}
\end{center}
\end{figure*}

The observed RM is a combination of an intrinsic RM of the background
sources plus an RM imposed by the foreground galaxy. The model of
observed Faraday rotations includes models of the random and regular
magnetic fields of the galaxy, the distribution of electron density
and the density of polarized background sources \cite{stepanov07}. The contribution of
the Galactic RM can be accounted properly from the density of RM
detected with the LOFAR (Low Frequency Array) or the SKA in a slightly
larger region surrounding each galaxy. The intrinsic RM of the
background sources will average out if the number of background
sources used for the analysis is large enough. \\
\emph{Models for the regular magnetic field in the disk}. Although the physics of dynamo
action still faces theoretical problems \cite{brand05}, the dynamo is
the only known mechanism which is able to generate large-scale {\em
coherent}\ (regular) magnetic fields of spiral shape. These coherent
fields can be represented as a superposition (spectrum) of modes with
different azimuthal symmetries. In a smooth, axisymmetric
gas disk the strongest mode is that with the azimuthal mode number
$m=0$ ({\em ASS}\ spiral field is observed in many nearby galaxies
\cite{berk03,gaensler05}; M\,31 in Fig.~\ref{fig:m31}),
followed by the weaker $m=1$ mode ({\em BSS}\ spiral fields are rare
\cite{beck05}), and $m=2$ ({\em QSS}) \cite{elstner92}. 
The two magnetic arms (located between the optical spiral arms) 
in NGC\,6946 \cite{beck07}, with the
field directed towards the galaxy's centre in both, is a signature of
superposed $m=0$ and $m=2$ modes. However, for many of the nearby
galaxies for which multi-frequency observations are available, the angular
resolutions and/or the signal-to-noise ratios are still insufficient to
reveal the dominating magnetic modes or their superpositions. We restrict
our analysis to the three lowest azimuthal modes of the toroidal
regular magnetic field: ASS, ($m=0$), BSS spiral ($m=1$), QSS spiral
($m=2$), and the superpositions ASS+BSS and ASS+QSS. All modes are
assumed to be symmetric with respect to the disk plane (S-modes). The
maximum regular field strength is assumed to be $5~\mu$G for all
modes, consistent with
typical values from observations \cite{beck05}. \\
\emph{Models for the regular magnetic field in the halo}. Strong
vertical fields (so-called X-shaped magnetic fields) extending into
the halo are recently observed in several edge-on galaxies such as e.g.
M\,104, NGC\,253, NGC\,891 and NGC\,5775
\cite{krause06,heesen05,krause07,soida05}. \\
\emph{Thermal electron density models.} We adopt a Gaussian
dependence of the electron density on galactic radius $r$ (in the
galactic plane) and on height $z$ above the disk midplane. The
electron density near the galaxy center, its radial scalelength and
vertical scalelength are taken to be $n_0=0.03$~cm$^{-3}$,
$r_0=10$~kpc and $h=1$~kpc, respectively. \\
\emph{Faraday rotation models.} The galactic magnetic field is
modeled as a superposition of a regular part with a simple azimuthal
symmetry $B_m$ and a random part $B_{turb}$ which describes the
contribution of large-scale galactic turbulence. The regular field
in the disk is parameterized by pitch angle and the intensity of the
corresponding azimuthal mode. The vertical field in the halo
increases with height and its direction is tilted at $45^\circ$ with
respect to the plane, as indicated by the magnetic field structure
observed in the halo of nearby galaxies. The Faraday rotation is
then modeled as the product of electron density and magnetic field
strength integrated along the line of
sight (see Fig.~\ref{fig:rm_model}). \\
\emph{The number density of polarized background sources} is
simulated using the number counts of polarized sources at 1.4\,GHz
\cite{taylor07} extrapolated to a polarized flux density limit
($\sim0.01$\,$\mu$Jy) reachable with the SKA.

\section{Recognition and reconstruction of regular field structures}
To recognize the large-scale regular component of the magnetic field
with a simple symmetry we apply the $\chi^2$ minimization between
the 'template' and the 'observed' RM patterns for ASS, BSS, and QSS
modes, or their superpositions. This method is also used to evaluate
the reliability of different modes. {\em Recognition} of single and
mixed modes of magnetic fields can be reliably performed from a
limited sample of RM measurements ($\ge20$) towards polarized
background sources. Higher modes (BSS and QSS) are easier to
recognize, i.e. they need less RM points, shorter observation time
and is less affected by the turbulent component. The dependence on
turbulence becomes dramatic for weakly inclined (almost face-on)
galaxies -- a reliable fitting requires a huge number of sources.

When the field structure is more complicated or the turbulent field
is stronger then the regular field, the \emph{reconstruction} method
should be applied. This does not need a ``precognition'' template,
but needs a higher density of RM sources and hence deeper
observations \cite{stepanov07}. To assess the number of RM points
required for a reliable reconstruction we compare the modeled and
reconstructed magnetic field profiles of the BSS-type galaxy.
The reconstruction method is superior for
strongly inclined galaxies (about $70^{\circ}$ and more) and is
successful if at least $20$ sources are available for one cut along
the {\it minor} galactic axis (or $\ge1200$ sources within the solid
angle covered by the galaxy).

\section{Perspectives for the SKA}
Recognition or reconstruction of regular field structures from RM
data of polarized background sources is a powerful tool for future
radio telescopes. Measuring RM at frequencies around 1~GHz with the
SKA, simple field structures can be recognized in galaxies up to
about 100\,Mpc distance and will allow to test dynamo against
primordial or other models of field origin for $\approx 60000$ disk
galaxies. For this the sensitivity of $0.015$~$\mu$Jy is needed, 
which can be achieved within 100~h observation time with the SKA. On
the other hand, the reconstruction of magnetic field structures of
strongly inclined spiral galaxies would require a sensitivity of the
SKA at 1.4~GHz of $\approx 0.5-5~\mu$Jy (or an integration time less
than one hour) for galaxies at distances of about one Mpc. The field
structures of $\approx 60$ galaxies at about 10~Mpc distance can be
reconstructed with tens to hundred hours of integration time.

A better knowledge of the \emph{slope of number counts} $\gamma$ of
polarized sources in the flux range accessible to the SKA is crucial
for observations of distant galaxies and of the required polarization
purity. Observing at larger distances $D$ requires to increase the
observing time $T$ according to $T\propto D^{4/\gamma}$ in order
to obtain the same number of sources per solid angle of a galaxy.
The recognition method raises requirements to the dynamic range of
the telescope for polarization measurements (polarization purity),
which may exceed 30~dB in the case of a flat slope of the number density
counts of polarized sources.

Depolarization effects within the background sources, in the
foreground of the galaxy or the Milky Way (in the case if the
angular extent of a source is larger than the angular turbulence
scale in the foreground medium) are yet uncertain. Both
depolarization effects, internal and external, depend on the angular
resolution and the observing frequency. No statistical data are
available yet. Future investigations in this direction may modify
the number of sources required for a reliable recognition.

As the RM errors are smaller at larger wavelengths the low-frequency
SKA array and low-frequency precursor telescopes like LOFAR would be
an ideal tool to study the weak magnetic fields in galaxies and
intergalactic space if background sources are still significantly
polarized at low frequencies.

%\bibliography{references}

\end{document}